\documentclass[10pt]{article}

\usepackage{graphicx}
\usepackage{bm}
\usepackage{amsfonts}
\usepackage{amsmath}
\usepackage{amssymb}
\usepackage{mathrsfs}

\usepackage{algorithm}
\usepackage{algorithmic}

\newtheorem{example}{Example}

\begin{document}

\title{Numerical Simulation of Two-phase Flow in Natural Fractured Reservoirs Using Dual Porosity Method on
Parallel Computers}

\author{Lihua Shen, Tao Cui, Hui Liu, Zhouyuan Zhu, He Zhong, Zhangxin Chen, \\
Bo Yang, Ruijian He, Huaqing Liu}

\date{}
\maketitle

\begin{abstract}
    The two-phase oil-water flow in natural fractured reservoirs and its numerical methods are
    introduced in this paper, where the fracture is modeled by the dual porosity method.
    Efficient numerical method, including the finite difference method,
    inexact Newton method nonlinear solver, CPR-FPF preconditioners for linear systems and effective
    decoupling method, are presented.
    Parallel computing techniques employed in simulation are also presented.
    Using those numerical methods and parallel techniques,
    a parallel reservoir simulator is developed, which is capable of simulating large-scale reservoir
    models. The numerical results show that our simulator is accurate and scalable compared to the commercial
    software, and the numerical methods are also effective.
\end{abstract}

{\bf \em keywords:} Reservoir simulation, Oil-water, Two-phase, Dual porosity, Fracture, Parallel computing

\section{Introduction}

Reservoir simuations are important tools for petroleum engineering, which have been applied to model
underground flows and interations between reservoirs and wells to predict the well
performance such as oil rates, water rates and bottom hole pressure. When a reservoir model is large enough,
it may take severals days or even longer for the simulator to complete a single model. Therefore, the numerical
methods and parallel techniques are essential to accelerate the simulations.

As a type of unconventional reservoirs, the reservoir with natural fractures and hydraulic fractures which are
commonly seen in tight and shale oil and gas reservoir, the dual-porosity/dual-permeability model and the
multiple iteration contiua (MINC) model\cite{pruess,wu-pruess} homogenize the fractures and use
superpositioned cells to represent the fractures and the matrix. The multiple continuum approaches have been
successfully employed in the unconventional reservoir problems. Wu et al.\cite{wu-qin,wu-di} regarded the
fractured vuggy rocks as a triple- or multiple-continuum medium with highly permeability and well-connected
fractures, low-permeabilty rock matrix and various-sized vugs.  Wu et al. also developed a hybrid
multiple-continuum-medium modeling appoach to describe different types of fractures including hydrautic
fractures, natural fracture network and micro fractures \cite{wu-li}. Jiang and Moinfar and their
collaborators designed explicit fracture
models coupled with MINC model to simulate the unconventional reservoirs \cite{moinfar,jiang}. In reference
\cite{wu-li-ding}, the authors presented a unified
framework model to incorporate the known mechanisms and processes including gas adsorption and desorption,
geomechanics effect, Klinkenberg or gas-slippage effect and non-Darcy flow. They also used a hybrid
fracture-modeling appoach to simulate the unconventional gas reservoirs. Jiang and Younis developed two
alternative hybrid approaches to capture the effects of the multiscale fracture systems by combining the
advantages of the multi-continuum and discrete-fracture/matrix representations  \cite{jiang-younis}.

The numerical methods for reservoir simulations have been explored by both mathematicians and engineers for
decades. Killough et al. \cite{killough} studied local refinement
techniques to improve the accuracy and reduce computational cost comparing to the global grid refinement.
Those local refinement techniques are useful for the complex models such as in-situ combustion.  Dogru and his
group \cite{dogru} developed parallel simulators using structured and unstructured grids to handle faults,
pinchouts, complex wells, polymer flooding in non-fractured and fractured reservoirs. Zhang et al.  developed
a general-purpose parallel platform for large-scale scientific applications which was designed using the
adaptive element methods and the adaptive finite volumn methods \cite{zhang,zhang-cui-liu}. It has
been applied to black oil simulation using discontinuous Galerkin methods\cite{wang-zhang-chen}. Chen et al.
studied finite element emtjods and finite different methods for black oil, compositional and thermal
models \cite{chen-huan-ma}. They studied Newton methods, linear solvers and preconditioners as well. 
Chen and his collaborators also developed a parallel platform to support the new-generation simulator
development, such as black oil, compositional, thermal, polymer flooding models 
\cite{para-frame,WLC-JCP,full-field}.  Wheeler studied
discretization methods, linear solvers, preconditioner techniques and developed a parallel black oil
simulator \cite{wheeler}. For the linear solver and preconditioners, many preconditioning methods have been
proposed and applied to reservoir simulations such as constrained pressure residual (CPR)
methods \cite{cpr,cao}, multi-stage methods \cite{dogru-wheeler}, multiple level preconditioners \cite{turkey},
fast auxiliary space preconditioners (FASP) \cite{auxiliary} and a family of parallel CPR-like
methods\cite{cpr}.

In this paper, we present the mathematical model of the dual-porosity reservoirs.
After that, we give the numerical methods used in our simulation inlcuding the nonlinear solvers,
the preconditioner and the linear solver and the parallel technique.
Finally, we show the numerical results obtained from our simulator.

\section{Mathematical Model}

Darcy's law describes the flow of a fluid through porous media, establishing the relationship
between the volumtric flow rate and the pressure gradient:
\begin{equation}
    Q = -\frac{KA\Delta p}{\mu L}
\end{equation}
where $K$ is the permeability of a given reservoir, $A$ is the area in the flow direction,
$\Delta p$ is the pressure difference, $\mu$ is the viscosity of the fluid and $L$ is the
length of the reservoir. In three dimensional space, its differential form is
\begin{equation}
    q =\frac{Q}{A}= -\frac{K\nabla p}{\mu }
\end{equation}
Combining Darcy's law and the mass conservation law for oil and water components, the
two-phase model is as follows:

\begin{equation}
    \left\{
        \begin{array}{ll}
            \frac{\partial}{\partial t}(\phi s_o\rho_o) &= \nabla \cdot (\frac{{\bm K}K_{ro}\rho_o}{\mu_o}\nabla \Phi_o) + q_o\\
            \frac{\partial}{\partial t}(\phi s_w \rho_w ) &= \nabla \cdot (\frac{{\bm K}K_{rw}\rho_w}{\mu_w} \nabla \Phi_w) + q_w\\
        \end{array}
        \right.
\end{equation}
where $\phi$ and ${\bm K}$ are the porosity and the permeabilities (in $x$-, $y$- and $z$-directions) respectively, 
$\Phi_\alpha$ is the potential, $s_\alpha$, $\mu_\alpha$, $K_{r\alpha}$ and $q_\alpha$
($\alpha = o, w$) are the saturation, viscosity, relative permeability and production/injection rate,
respectively, which satisfy the following constraints:
\begin{equation}
    \left\{
        \begin{array}{l}
            \Phi_\alpha = p_\alpha - \rho_\alpha g z,\\
            s_o+s_w = 1,\\
            p_w = p_o - p_{cow}(s_w),\\
        \end{array}
        \right.
\end{equation}
where  $p_\alpha$ and $\rho_\alpha$ are the pressure and the mass density of phase $\alpha$($\alpha=o,~w$). $p_{cow}$
is the capillary pressure between oil phase and water phase, $g$ is the gravitational constant and $z$
is the reservoir depth.

\begin{figure}[!ht]\centering
    \includegraphics[scale = 0.2]{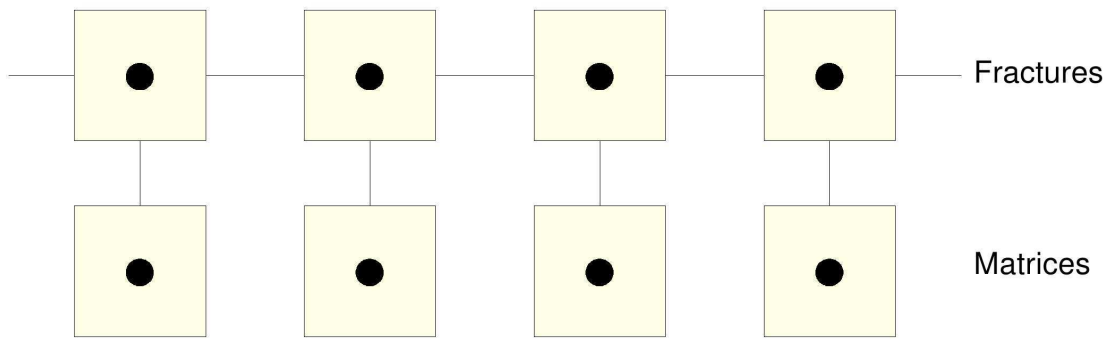}
    \caption{The flow commmunication for dual porosity model \cite{cmg}}
    \label{matrix-fracture}
\end{figure}

For the dual-porosity model, fluids can flow between matrix and fracture, as well as fracture and fracture,
but there is no flow between matrices. There flow directions are described by Fig \ref{matrix-fracture}.
The following equations are used to describe this model:
\begin{equation}\label{prob}
    \left\{
        \begin{array}{l}
            \frac{\partial}{\partial t}({\phi_f s_{o,f} \rho_{o,f}}) = \nabla \cdot (\frac{{\bm K}_fK_{ro_f}
            \rho_{o,f}}{\mu_{o,f}}\nabla \Phi_{o,f})- q_{o,mf}+ q_o,\\
            \frac{\partial}{\partial t}({\phi_m s_{o,m} \rho_{o,f}}) = q_{o,mf},\\
            \frac{\partial}{\partial t}({\phi_f s_{w,f} \rho_{w,f}}) = \nabla \cdot (\frac{{\bm K}_fK_{rw_f}
            \rho_{w,f}}{\mu_{w,f}b_{w,f}}\nabla \Phi_{w,f})-q_{w,mf}+ q_w,\\
            \frac{\partial}{\partial t}({\phi_m s_{w,m} \rho_{w,m}}) = q_{w,mf}
        \end{array}
        \right.
\end{equation}
where $(\cdot)_f$ denotes the variables for the fracture and $(\cdot)_m$ for the matrix.
Note that here $q_{\alpha,mf} = \omega (\frac{K_{r\alpha} \rho_{\alpha,t}}{\mu_\alpha})_t(p_f - p_m)$ is a source
term which represents the net addition of the fluid to the fracture from the matrix with $\omega$ the shape
factor. In our simulator, the shape factor we use is based on the work of Warren and Root\cite{warren-root}
and Gilman and Kazemi\cite{gilman-kazemi}.  Note that $(\cdot)_t$ denoted the value at $t
= m$ or $t = f$ depending on which is the upsteam. The well rate at a perforation, $i$, is modelled by the sink-source method,
\begin{equation}
    q_{\alpha,i} = W_i \frac{{\bm K}_fK_{r\alpha_f} \rho_{\alpha,f}}{\mu_{\alpha,f}}(p_b - p), \quad \alpha = o,~w
\end{equation}
where $W_i$ is the well index at the perforation, $p_b$ is the bottom-hole pressure of the well at the perforation, $p$ is the
block pressure from oil phase at the perforation.
For a producer or an injection well, its well rate is calculated as,
\begin{equation}
    q_\alpha = \sum_{i}q_{\alpha,i},  \quad \alpha = o,~w.
\end{equation}
The well index is calculated as follows:
\begin{equation}
    W_i = 2\pi* k_h*w_{frac}/(\ln(r_e/r_w)+s).
\end{equation}
Here $k_h$ is a given value and $w_{frac}$ is the facture between 0.0 and 1.0 specifying the fraction of a circle
that the well models, $r_w$ is the wellbore radius, $s$ is a real number specifying the well skin factor, 
$r_e$ is the well effective radius calculated from 
\begin{equation}
    r_e = w_{g}*\sqrt{A_{r}/(\pi*w_{frac})} 
\end{equation}
with $w_{g}$ the geometric factor for the well element, $A_{r}$ the area perpendicular to the 
wellbore (e.g., $D_x*D_y$ for a vertical well) and $w_{frac}$ as same above.
The other option for the well effective radius is the following Peaceman's form \cite{peaceman}
\begin{equation}\displaystyle
    \left\{
        \begin{array}{l}
            r_{e,x} = 0.28\frac{\displaystyle[D_x^2(\frac{k_y}{k_x})^{1/2} + D_y^2(\frac{k_x}{k_y})^{1/2}]^{\frac{1}{2}}}{\displaystyle(\frac{k_y}{k_x})^{1/4}+ (\frac{k_x}{k_y})^{1/4}},\\
            r_{e,y} = 0.28\frac{\displaystyle[D_x^2(\frac{k_z}{k_x})^{1/2} + D_z^2(\frac{k_x}{k_z})^{1/2}]^{\frac{1}{2}}}{\displaystyle(\frac{k_z}{k_x})^{1/4}+ (\frac{k_x}{k_z})^{1/4}},\\
            r_{e,z} = 0.28\frac{\displaystyle[D_y^2(\frac{k_z}{k_y})^{1/2} + D_z^2(\frac{k_y}{k_z})^{1/2}]^{\frac{1}{2}}}{(\displaystyle\frac{k_z}{k_y})^{1/4}+ (\frac{k_y}{k_z})^{1/4}},\\
        \end{array}
        \right.
\end{equation}
where $r_{e,x}$, $r_{e,y}$ and $r_{e,z}$ are the well effective radius in $x$, $y$ and $z$ direction respectively, 
$D_x$, $D_y$ and $D_z$ are the block sizes in $x$, $y$ and $z$ direction respectively, $k_x$, $k_y$ and $k_z$
are the permeabilities in $x$, $y$ and $z$ direction respectively.

\section{Numerical Methods and Parallel Computing}

In this paper, the fully implicit method (FIM) is applie, and the oil phase pressure $p$, water saturation
$s_w$ and the well bottom hole pressure $p_b$ are unknowns. For the time differential and the space
differential term, we use the backward Euler scheme and cell-centered finite difference method as
discretization methods.

The system is highly nonlinear,
\begin{equation}\label{nonlinear}
    F(x) = 0
\end{equation}
where $F$ is a nonlinear map from $\mathcal {R}^N$ to $\mathcal {R}^N$ with $N =  2
\times n + \tau$. Here $n$ is the number of grid blocks and $\tau$ is the well size.
In this nonlinear equation, the properties related to the saturation are strongly nonlinear
while the properties related to the pressure are weakly nonlinear.
In our simulation, we use Newton method (or inexact Newton method) to solve it.

\subsection{Nonlinear Solver}
The inexact Newton method can be regarded as an extension of the standard Newton method. Let $A$ be the Jacobian
matrix of $F(x)$ and $y$ is the correction between the last step approximate solution and the current step one.
Its algorithm \cite{chen-gewecke} is as follows:

\begin{algorithm}[!htb]
\caption{The Inexact Newton Method}
\label{inewton-alg}
\begin{algorithmic}[1]
\STATE Given an initial guess $x^0$ and a termination tolerance $\epsilon$, let $l = 0$
 and assemble the right-hand side $b$.
\WHILE{$\left\|b \right\| \ge \epsilon$}
\STATE Compute the Jacobian matrix $A$.
\STATE Determine $\eta_l$.
\STATE Find a solution $y$ such that
    \begin{equation}
    \label{inexact-newton-alg}
    \left\| A y - b \right\| \leq \eta_l \left\| b \right\|,
    \end{equation}
\STATE    Let $l = l+1$ and $x^l = x^{l-1} + y$.
\STATE    Compute the new right-hand side $b$
\ENDWHILE
\STATE $x^* = x^l$ is the approximate solution of the nonlinear system, $F(x) = 0$.
\end{algorithmic}
\end{algorithm}

We can see that the algorithm is the same as Newton method except the choice 
of the tolerance $\theta$. The standard Newton method usually applies a small constant such as $1.0e-6$, while
the {\bf Algorithm 1} uses adaptive $\theta_l$ tolerance to avoid over solution. Common methods are defined
as follows \cite{chen-gewecke}:
\begin{equation}
    \theta_l =\left\{ 
    \begin{array}{l}
        \displaystyle\frac{\|b(x^l) - r^{l-1}\|}{\|b(x^{l-1})\|},\\
        \displaystyle\frac{\|b(x^l)\| - \|r^{l-1}\|}{\|b(x^{l-1})\|},\\
        \displaystyle\gamma (\frac{\|b(x^l)\|}{\|b(x^{l-1})\|})^\beta, \quad \gamma\in [0,~1], \quad \beta\in (1,2]
    \end{array}
    \right.
\end{equation}
where $r^l$ is the residual of the $l$-th iteration.

\subsection{Preconditioner and Linear Solver}

For each Newton iteration, we need to solve a linear system $Ax = b$ which is very time
comsuming. To improve the efficiency, we choose a proper preconditioner, CPR-FPF method \cite{cpr}, for the
model. Each grid block has four unknowns, pressures ($p_f$ and $p_m$) and saturations ($s_{w,f}$ and
$s_{w,m}$) for matrix and fracture, and four
equations. In our linear system, each unknown and each row are arranged cell by cell, in this case, the
Jacobian matrix is a block matrix. It is also well-known that block matrix has better convergence that
point-wise matrix.

Before solving the linear system $Ax=b$, a decoupling operation $D$ is applied, and an equivalent
system is obtained:
\begin{equation}\label{decoupled-eq}
    D^{-1}Ax = D^{-1} b.
\end{equation}
A proper decoupling method can improve linear solver dramatically. Many 
decoupling strategies have been proposed, such as Quasi-IMPES\cite{quasi-impes} method and ABF\cite{abf}
method. In this paper, a modified Gauss-Jordan elimination method is applied.

\subsection{Parallel Computing}

The simulator is based on our parallel platform PRSI \cite{para-frame}, which is developed using C language
and MPI (Message Passing Interface). The platform has implemented many modules, such as
grid generation, load balancing, well management, parallel input and output, distributed matrix and vector, linear solver
and preconditioner, communication management and visualization.
Based on the platform, physical modules, such as rock properties and rock-fluid properties, are implemented.
More details can be read from reference \cite{para-frame}.

\section{Numerical Results}

\begin{example}
    \label{ex1}
    The grid dimension is $10\times 10\times 1$ with sizes 102.04 ft. in $x$ and $y$ directions
    and  100.0 ft. in $z$ direction from top to bottom. The depth of the top layer center is 2000.0 ft.
    The permeabilities for the matrix in $x$, $y$ and $z$ directions are 100, 100, 10 mD respectively.
    The permeability for the fracture in $x$, $y$ and $z$ directions are 395.85 mD.
    The reference porosity for the matrix is 0.1392. The reference porosity for the fracture is 0.039585.
    The rock compressibilities for the matrix and fracture are both 3e-06 (1/psi). The reference pressure
    is 15.0 psi for both the matrix and the fracture.
    Component properties: densities of oil and water are 58.0 lbm/ft3 and 62.4 lbm/ft3 respectively.
    The reference pressure is 15 psi at which the oil formation volume factor is 1.036 RB/STB and the oil viscosity is 40.0 cp. 
    The oil compressibility is const 1.313e-5 l/psi.
    The initial conditions are as follows: initial pressure for the matrix is 2000 psi, initial pressure for the fracture is 1980 psi,
    and initial water saturations are 0.08 and 0.01 in matrix and fracture respectively.
    There are one injection well and two production wells. All of them are vertical.
    Injection well has maximum water injection rate 500.0 bbl/day, maximum bottom
    hole pressure 5.0e+4 psi, well index 200.0 with perforation at cell [5 1 1].
    Both of the production well has maximum oil production rate 300.0 STB/day, minimum bottom hole pressure 15 psi
    with well radius 0.25ft. The perforation of Produer 1 is at cell [1 10 1] while the perforation of Produer 2 is at cell [10 10 1].
    The simulation period is 800 days.
\end{example}

\begin{figure}[!ht]
    \centering
    \includegraphics[width = 0.5\textwidth]{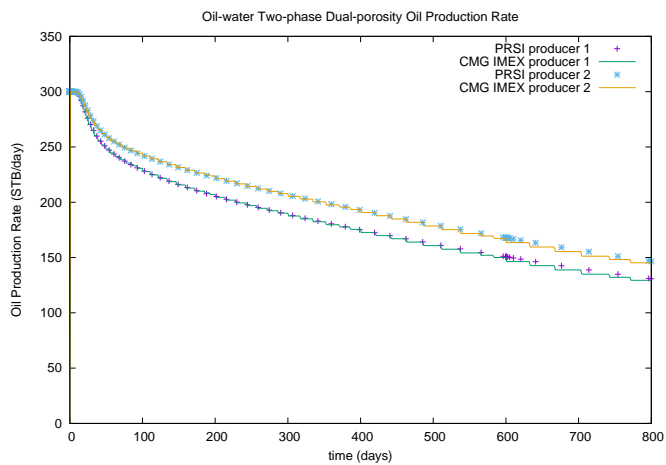}
    \caption{Example \ref{ex1}: Oil production rate (unit: STB/day).}
    \label{fig-dp-oil-rate}
\end{figure}

\begin{figure}[!ht]
    \centering
    \includegraphics[width = 0.5\textwidth]{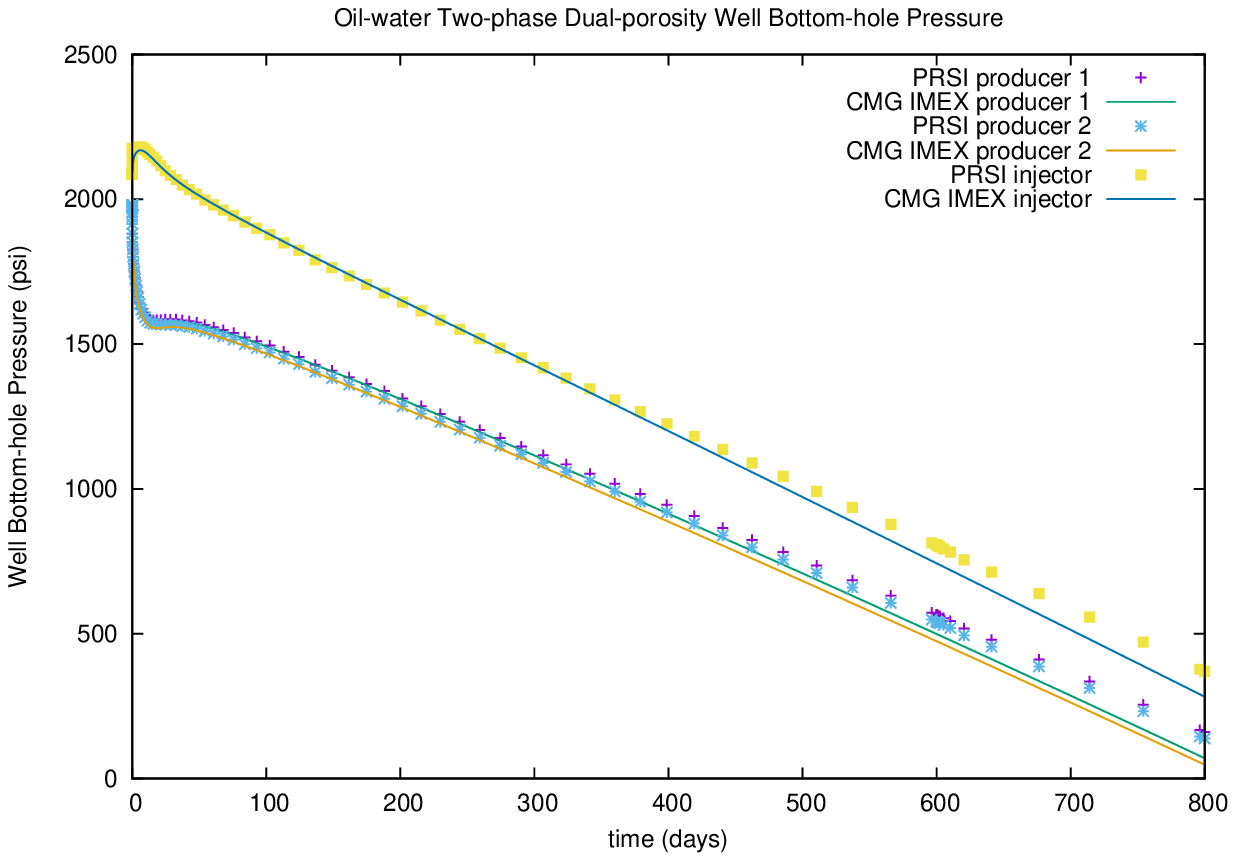}
    \caption{Example \ref{ex1}: Well bottom-hole pressure (pressure unit: psi).}
    \label{fig-dp-bhp}
\end{figure}

\begin{figure}[!ht]\centering
    \includegraphics[width = 0.5\textwidth]{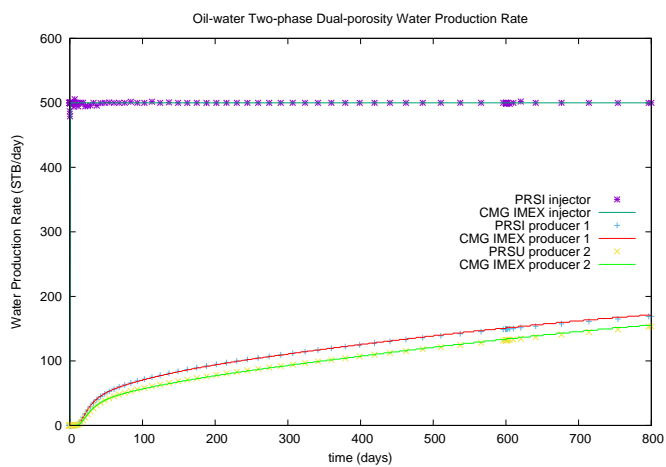}
    \caption{Example \ref{ex1}: Water rate (unit: STB/day).}
    \label{fig-dp-water-rate}
\end{figure}

The results of oil production rate, bottom-hole-pressure and water rate are shown in 
Fig. \ref{fig-dp-oil-rate}, \ref{fig-dp-bhp}, \ref{fig-dp-water-rate}, from which we can see that the results 
from our simulator and from CMG IMEX match very well. This proves our methods and implementation are
correct.

\begin{example}
    \label{ex2}
    The grid dimension is $10\times 10 \times 3$ with mesh size 102 ft. in $x$ and $y$ directions
    and  100.0 ft. in $z$ direction from top to bottom. The depth of the top layer
    center is 2000.0 ft.
    The permeabilities for the matrix in $x$, $y$ and $z$ directions are 100, 100, 100 mD respectively.
    The permeability for the fracture in $x$, $y$ and $z$ directions are 395.85 mD.
    The reference porosity for the matrix is 0.1392. The reference porosity for the fracture is 0.039585.
    The rock compressibilities for the matrix and fracture are both 3e-06 (1/psi). The reference pressure
    is 15.0 for both the matrix and the fracture.
    Component properties: densities of oil and water are 58.0 lbm/ft3 and 62.4 lbm/ft3 respectively.
    PVT: the reference pressure is 15 psi at which the oil formation volume factor is 1.036 RB/STB and the oil viscosity is 40.0 cp.
    The oil compressibility is const 1.313e-5 l/psi.
    The initial conditions are as follows: initial pressure for the matrix is 800 psi, initial pressure for the fracture is 500 psi,
    and initial water saturations are 0.08 and 0.01 in matrix and fracture respectively.
    There are one injection well and two production wells. All of them are vertical.
    Injection well has maximum water injection rate 200.0 bbl/day, maximum bottom
    hole pressure 5.0e+4 psi, well index 200.0 with perforation
    at cell [5 5 1].
    Both of the production well has maximum oil production rate 500.0 STB/day, minimum 
    bottom hole pressure 15 psi with well radius 0.25ft. The perforation of Produer 1 is at 
    cell [1 1 1] while the perforation of Produer 2 is at cell [10 10 1].
    The simulation period is 1600 days.
\end{example}

\begin{figure}[!ht]
    \centering
    \includegraphics[width = 0.5\textwidth]{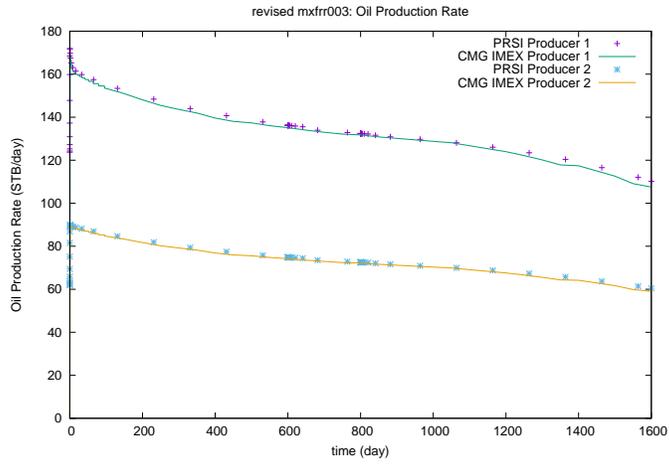}
    \caption{Example \ref{ex2}: Oil production rate (unit: STB/day).}
    \label{fig-mxfrr3-revised-oil-rate}
\end{figure}

\begin{figure}[!ht]
    \centering
    \includegraphics[width = 0.5\textwidth]{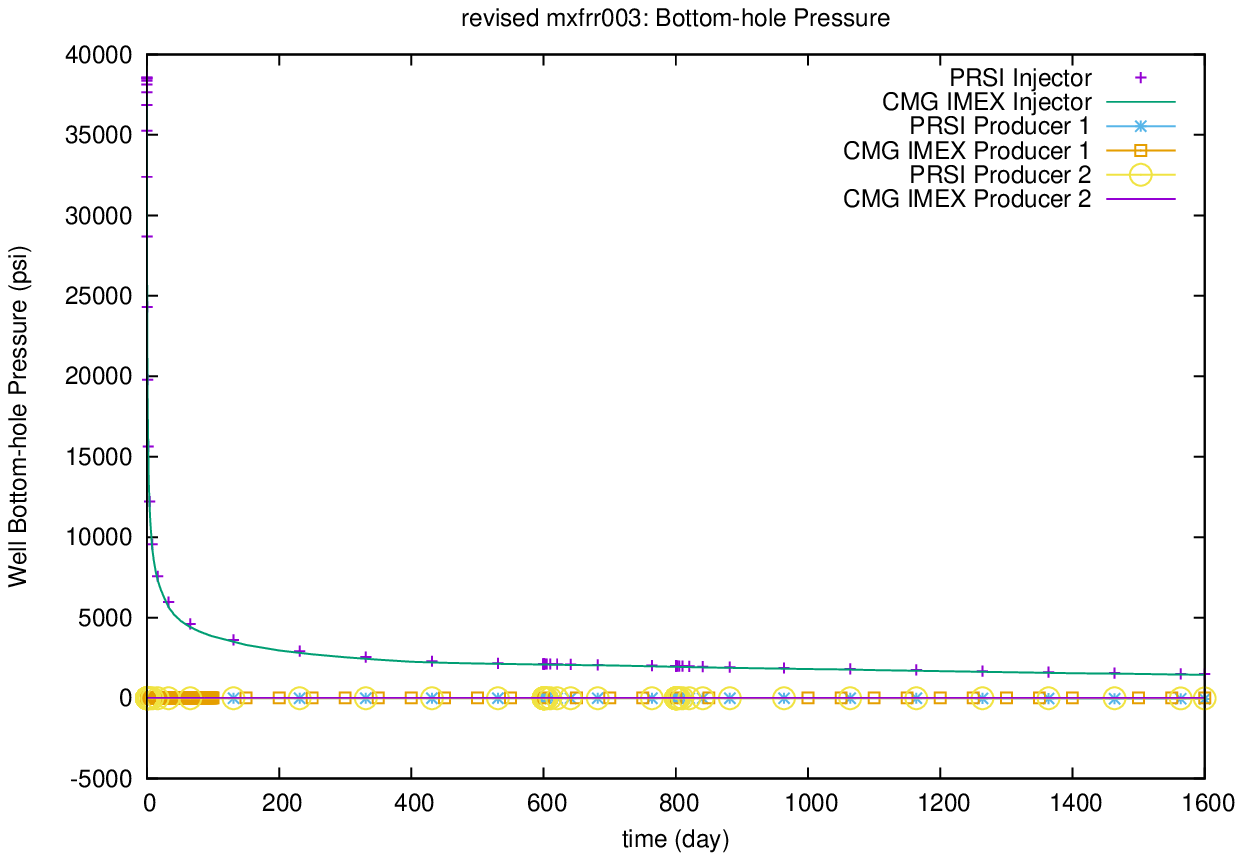}
    \caption{Example \ref{ex2}: Well bottom-hole pressure (pressure unit: psi).}
    \label{fig-mxfrr3-revised-bhp}
\end{figure}

\begin{figure}[!ht]
    \centering
    \includegraphics[width = 0.5\textwidth]{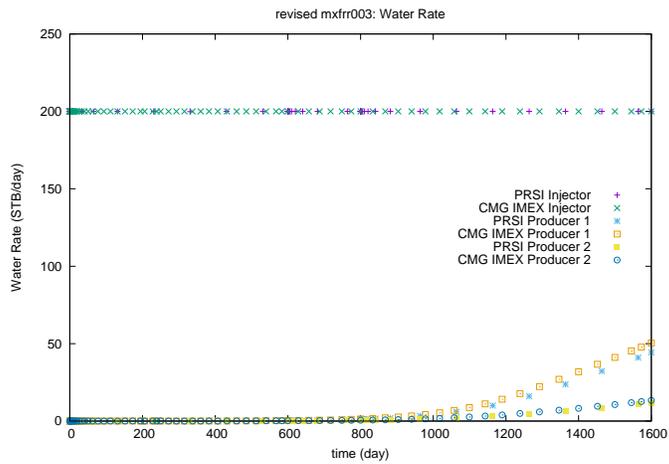}
    \caption{Example \ref{ex2}: Water rate (unit: STB/day).}
    \label{fig-mxfrr3-revised-water-rate}
\end{figure}

The results of oil production rate, bottom-hole-pressure and water rate are shown in 
Fig. \ref{fig-mxfrr3-revised-oil-rate}, \ref{fig-mxfrr3-revised-bhp}, \ref{fig-mxfrr3-revised-water-rate}.
Again, these figures show that our simulator match CMG simulator.

\begin{example}
    \label{ex3}
    This example tests the scalability of our two-phase dual porosity simulator by computing Example 1 with
    grid dimension $500\times 500\times 50$.
\end{example}

\begin{figure}[!ht]\centering
    \includegraphics[width = 0.5\textwidth]{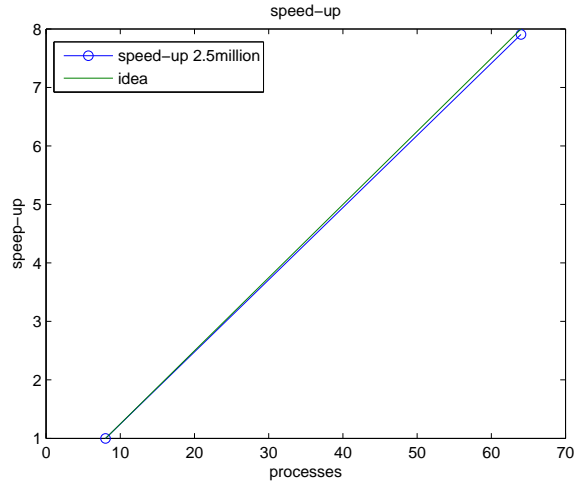}
    \caption{Example \ref{ex3}: speed-up.}
    \label{fig-scale}.
\end{figure}

As we introducerd before, we employ the GMRES linear solver and CPR preconditioner. Table \ref{tab-scale}
shows the numerical summaries, which show that our numerical methods are stable when increasing CPU cores. 
Figure \ref{fig-scale} presents the scalability of our simulator, which demonstrates that the simulator
and parallel implementation are scalable.

\begin{table}[!ht]
    \centering
    \begin{tabular}{|c|c|c|c|c|}
        \hline
        \#MPIs & 8 & 64 \\
        \#Time steps & 68 & 68  \\
        \#Newton iterations& 273 & 267 \\
        \#total linear iterations & 641& 623 \\
        \#total running time(s) &20009.71 & 2531.91 \\
        \hline
    \end{tabular}\label{tab-scale}

\end{table}

\section{Conclusion}

This paper presents our work on development of two-phase oil-water simulator for natural fractured
reservoirs using dual porosity method. Effective numerical methods and parallel computing techniques
are introduced.
From the numerical experiments, we can see that our results match commercial simulator, CMG IMEX,
and the simulatro has good parallel scalability and is capable of handling large scale reservoir simulation
problems.

\section{Acknowledgements}
This work is partially supported by Department of Chemical Petroleum Engineering, 
University of Calgary, NSERC, AIEES, Foundation CMG, AITF iCore, IBM Thomas J. Watson 
Research Center, Frank and Sarah Meyer FCMG Collaboration Center, 
WestGrid (www.westgrid.ca), 
SciNet (www.scinetpc.ca) and Compute Canada Calcul Canada (www.computecanada.ca).


\begin{thebibliography}{10}

\bibitem{dogru-wheeler} T. Al-Shaalan, H. Klie, A. Dogru, and M. Wheeler, Studies of robust two stage preconditioners
    for the solution of fully implicit multiphase flow problems, SPE Reservoir Simulation Symposium, 2009.

\bibitem{abf} R. Bank, T. Chan, W. Coughran Jr., and R. Smith, The Alternate-Block-Factorization
    procedure for systems of partial differential equations, BIT Numerical Mathematics, 29(4), 1989, 938-954.

\bibitem{cao} H. Cao, T. Schlumberger, A. Hamdi, J. Wallis, and H. Yardumian, Parallel scalable unstructured
    CPR-type linear solver for reservoir simulation, SPE Annual Technical Conference and Exhibition, 2005.

\bibitem{ras} X. Cai and M. Sarkis, A restricted additive Schwarz preconditioner for general sparse
    linear systems, SIAM Journal on Scientific Computing, 21(2), 1999, 792-797.


\bibitem{chen-gewecke} T. Chen, N. Gewecke, Z. Li, A. Rubiano, R. Shuttleworth B. Yong and X. Zhong, Fast Comutational
    Methods for Reservoir Flow Models, 2009.


\bibitem{chen-huan-ma} Z. Chen, G. Huan and Y. Ma, Computational methods for multiphase flows in porous media, Vol. 2, Siam, 2006.

\bibitem{dogru} A. Dogru, L. Fung, U. Middya, T. Al-Shaalan, and J. Pita, A next-generation parallel
    reservoir simulation for the giant reservoirs, SPE/EAGE Reservoir Charaterization \& Simulation Conference, 2009.

\bibitem{gilman-kazemi} J.R. Gilman and H. Kazemi, Improvements in Simulation of Naturally Fractured Reservoirs, SPE-10511-PA, Vol. 23(04), 1983, 695-707.

\bibitem{auxiliary} X. Hu, W. Liu, G. Qin, J. Xu, and Z. Zhang, Development of a fast auxiliary subspace pre-conditoner
    for numerical reservoir simulators, SPE Reservoir Charaterisation and Simulation Conference
    and Exhibition, 2011.


\bibitem{jiang} J. Jiang, Y. Shao, R. M. Younis, et al., Development of a multi-continuum
    multi-component model for enhanced gas recovery and CO2 storage in fractured shale gas reservoirs, SPE
    Improved Oil Recovery Symposium, Society of Petroleum Engineers, 2014.

\bibitem{jiang-younis} J. Jiang, R. Younis, et al., Hybrid coupled discrete-fracture/matrix and multi-contimuum
    models for unconvenctional-reservoir simulation, SPE Journal 21(03), 2016 1009-1027.

\bibitem{killough} J. Killough, D. Camilleri, B. Darlow, and J. Foster, Parallel reservoid simulation based on local grid refinement. SPE-37978,
    SPE Reservoir Simulation Symposium., Dallas, 1997.


\bibitem{quasi-impes} S. Lacroix, Y. Vassilevski, and M. Wheeler, Decoupling preconditioners in
    the implicit parallel accurate reservoir simulation (IPARS), Numerical linear algebra with applications,
    8(8), 2001, 537-549.

\bibitem{cpr} H. Liu, K. Wang, and Z. Chen, A family of constrained pressure residual preconditioners for parallel
    reservoir simulations, Numerical Linear Algebra with Applications, Vol. 23(1), 2016, 120-146.

\bibitem{hsfc} H. Liu, Dynamic load balancing on adaptive unstructured meshes, 10th IEEE International Conference
    on High Performace Computing and Communications, 2008.


\bibitem{para-frame} H. Liu, K. Wang, Z. Chen, K. Jordan, J. Luo, and H. Deng, A parallel framework
    for  reservoir simulators on distributed-memory supercomputers, SPE-17645-MS, SPE/IATMI Asia Pacific
    Oil \& Gas Conference and Exhibition, 20-22 October, Nusa Dua, Bali, Indonesia, 2015.

\bibitem{moinfar} A. Moinfar, A. Varavei, K. Sepehrnoori, R. T. Johns, et al.,
    Development of a coupled dual continuum and discrete fracture model for the simulation
    of unconventional reservoirs, SPE Reservoir Simulation Symposium, Society of Petroleum Engineers, 2013.

\bibitem{peaceman}  Peaceman, D. W., Interpretation of Well-Block Pressures in Numerical Reservoir Simulation with Non-Square
    Grid Blocks and Anisotropic Permeability, SPE Journal, 23(3), 1983, 531-543.


\bibitem{pruess} K. Pruess, et al. A practical method for modeling fluid and heat flow in fractured porous
    media, Society of Petroleum Engineers Journal, Vol. 25(01), 14-26 1985.


\bibitem{wallis} J. Wallis, R. Kendall, and T. Little, Constrained residual acceleration of conjugate
    residual methods, SPE Reservoir Simulation Symposium, 1985.

\bibitem{wang-zhang-chen} K. Wang, L. Zhang, Z. Chen, Development of discontinuous Galerkin methods and a
    parallel simulator for reservoir simulation, SPE-176168-MS, SPE/IATMI Asia Pacific
    Oil \& Gas Conference and Exhibition, 20-22 October, Nusa Dua, Bali, Indonesia, 2015.


\bibitem{WLC-JCP} K. Wang, H. Liu, and Z. Chen, A scalable parallel black oil simulator on distributed memory
    parallel computers, Journal of Computational Physics, Vol. 301, 19-34.

\bibitem{full-field} K. Wang, H. Liu, J. Luo, and Z. Chen, Parallel simulation of full-field polymer flooding, The
    2nd IEEE International Conference on High Performance and Smart Computing, 2016.


\bibitem{turkey} B. Wang, S. Wu, Q. Li, H. Li, C. Zhang, and J. Xu, A multilevel preconditioner and its
    shared memory implementation for new generation reservoir simulation, SPE-172988-MS, SPE Large Scale Computing
    and Big Data Challenges in Reservoir Simulation Conference and Exibition, 15-17 September, Istanbul, Turkey, 2014.

\bibitem{warren-root} J.E. Warren and P.J. Root, The Behavior of Naturally Fractured Reservoirs, SPE-426-PA, Vol. 3(02), 1963.  245-255.

\bibitem{wheeler} M. Wheeler, Advanced techniques and algorithms for reservoir simulation, II: The multiblock
    approch in the integrated parallel accurate reservoir simulation (IPARS), the IMA Volumes in
    Mathematics and its Applications, Springer New York, 9-19, 2002.

\bibitem{wu-li} Y.S Wu, N. Li, C. Wang, Q. Ran, J. Li, J. Yuan, et al., A multiple-continuum
    model for simulation of gas production from shale gas reservoirs, SPE Reservoir Characterization and
    Simulation Conference and Exhibition, Society of Petroleum Engineers, 2013.

\bibitem{wu-qin} Y.S. Wu, G. Qin, R.E.Ewing, Y. Efendiev, Z. Kang, Y. Ren, et al.,
    A multiple-continuum approach for modeling multiphase flow in naturally fractured vuggy petroleum reservoirs:
    International Oil \& Gas Conference and Exhibition in China, Society of Petroleum Engineers, 2006.

\bibitem{wu-di} Y.S. Wu, Y. Di, Z. Kang, P. Fakcharoenphol, A multiple-continuum
    model for simulating single-phase and multiphase flow in naturally fractured vuggy
    reservoirs, Journal of Petroleum Science and Engineering, 78(1), 2011, 13-22.

\bibitem{wu-pruess}
    Y.S. Wu, K. Pruess, et al., A multiple-porosity method for simulation of
    naturally fractured petroleum reservoir, SPE Reservoir Engineering 3(01), 1998, 327-336.

\bibitem{wu-li-ding}
    Y.S. Wu, J. Li, D. Ding, C. Wang, Y. Di, et al., A generalized framework
    model for the simulation of gas production in unconventional gas reservoirs, SPE Journal 19(05), 2014, 845-857.

\bibitem{zhang} L. Zhang, A parallel algorithm for adaptive local refinement of
    tetrahedral meshes using bisection, Numer. Math.: Theory, Methods and applications, Vol. 2, 65-89 2009.

\bibitem{zhang-cui-liu} L. Zhang, T. Cui and H. Hui, A set of symmetric quadrature
    rules on triangles and tetrahedra, J. Comput. Math. 2009, 27(1), 89-96.

\bibitem{cmg} CMG Ltd, IMEX User Guid, Version 2014, 2014. 

\end{thebibliography}
\end{document}